\documentclass[twocolumn]{aastex631}

\renewcommand{\edit}[1]{}

\begin{document}

\title{Field Line Universal relaXer (FLUX): \\ A Fluxon Approach to Coronal Magnetic Field Modeling}
\correspondingauthor{Chris Lowder}
\email{chris.lowder@swri.org}

\author[0000-0001-8318-8229]{Chris Lowder}
\affiliation{Southwest Research Institute, Boulder, CO 80302, USA}

\author[0000-0003-0021-9056]{Chris Gilly}
\affiliation{Southwest Research Institute, Boulder, CO 80302, USA}

\author[0000-0002-7164-2786]{Craig DeForest}
\affiliation{Southwest Research Institute, Boulder, CO 80302, USA}

\begin{abstract}

We describe a novel method for modeling the global, steady solar wind using photospheric magnetic fields as a driving boundary condition. Prior wind models in this class include both rapid heuristic methods that use potential field extrapolation and variants thereof, trading rigor for computation speed, and detailed 3D magnetohydrodynamic (MHD) models that attempt to simulate the entire solar corona with a degree of physical rigor, but require large amounts of computation. The Field Line Universal relaXer (FLUX), an open-source numerical code which implements the `fluxon' semi-lagrangian approach to MHD modeling, provides an intermediate approach between these two general classes. In particular, the fluxon approach to MHD describes the magnetic field through discrete analogues of magnetic field lines, relaxing these structures to a stationary state of force balance. In this work we introduce a one-dimensional solar wind solution along each fieldline, providing an ensemble of solutions that are interpolated back onto a uniform grid at an outer boundary surface. This provides advantages in physical rigor over heuristic semi-analytic techniques, and in computational efficiency over full 3D MHD techniques. Here we describe the underlying methodology and the FLUXPipe modeling pipeline process.

\end{abstract}

\section{Introduction}


The solar wind plays a primary role in the space weather environment at Earth. Further, the wind is thought to be driven by boundary conditions low down: magnetic conditions near the photosphere and physical conditions in the solar corona \citep[eg.][]{1958ApJ...128..664P, 1973NASSP.303.....G, 1977RvGSP..15..257Z, 2009LRSP....6....3C}. The magnetic field shapes the solar wind both as a source of energy (and, indirectly, mass from the solar photosphere) and as a conduit that affects the accelerating flow through the spatial relationships between local flux tube geometry and location of energy and momentum deposition \citep[eg.][]{1993ApJ...410L.123W, 1998SoPh..180..231S}.


Current models of global solar wind flow fall into two major classes: (1) heuristic models such as WSA \edit1{\citep{1990ApJ...355..726W, 2000JGR...10510465A, 2003AIPC..679..190A, 2004JASTP..66.1295A}} or WSA/CSSS \citep{2014ApJ...782L..22P} that rely on analytic extrapolation of field geometry, coupled with empirical heuristics to describe the relationship between field geometry and wind flow; and (2) 3D magnetohydrodynamic models that attempt to simulate the corona directly, using resistive MHD and models of the underlying physics \edit1{\citep{2010ApJ...712.1219D, 2009ApJ...691..640R, 2012LRSP....9....6M}}. The former are very fast to implement, but are very limited in their ability to capture the coronal geometry and solar wind behavior. The latter \edit1{more directly model the underlying physics of the solar corona, however} are famously very difficult and slow to implement, \edit1{and they} also suffer from effects such as numerical resistivity that diffuses the sharp topological boundaries observed in the physical corona.


The Field Line Universal relaXer (FLUX) simulation code models the solar corona as a collection of discrete analogue magnetic domains, via a quasi-Lagrangian grid of discrete field lines (`fluxons'). Each of these fluxon structures carries a finite quantity of magnetic flux, interacting with neighboring fluxons and relaxing in time. This approach is computationally efficient, scales easily, and avoids certain grid-based numerical issues \edit1{- falling between the two classes of models previously described}. This provides a unique and multi-functional tool for coronal modeling. Each fluxon provides an ideal framework for one-dimensional solar wind modeling, which in concert provides full solar wind mapping to the edge of the corona.

The fluxon approach has been used primarily to study magnetic stability and the importance of reconnection to magnetic eruption processes such as CMEs \citep{2009ApJ...693.1431R,2010ApJ...715.1556R}, in the `zero-beta' domain with no included plasma. Recently, we have adapted the FLUX code to support steady solar wind solutions and to interact with other modeling codes in a hybrid modeling framework. The techniques we adopted are of interest because of the unique challenges of the fluxon approach to modeling, and in the present article we describe those techniques as a prelude to future work comparing the fluxon framework to other solar wind models.

The fluxon method was originally described by \edit1{\cite{2007JASTP..69..116D}}; here we describe the current implementation and its application to solar wind modeling, including adaptations to both the code and the underlying force laws and method. In \S\ref{sec:methodology} we describe the methodology behind this work, including the building blocks and mechanics of the fluxon model (\S\ref{sec:cormodeling}) and methods of solar wind modeling (\S\ref{sec:windmodeling}). The encapsulation of an automated framework for fluxon modeling of the solar wind is described in \S\ref{sec:fluxpipe}. Results from the model are described in \S\ref{sec:results}. A summary of results and discussion are provided in \S\ref{sec:discussion}, with final thoughts and conclusions in \S\ref{sec:conclusions}.

\section{Methodology} \label{sec:methodology}

\subsection{Coronal modeling} \label{sec:cormodeling}

\begin{figure}[!htbp]
\plotone{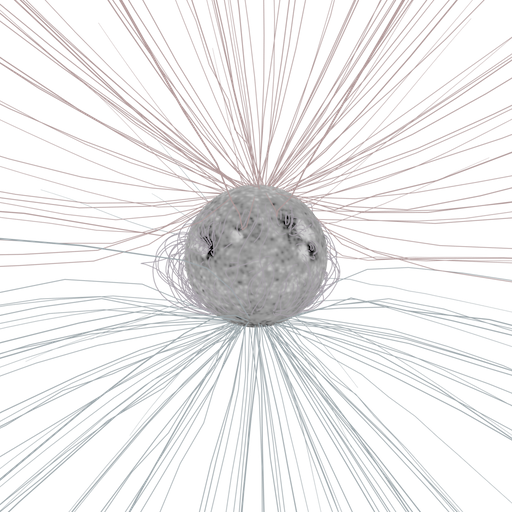}
\caption{Relaxed geometry from the FLUX model. A synthetic map of $B_r$ is displayed in \edit1{gray}, with fluxons color mapped by radial extent.}
\label{fig:fluxonmodel}
\end{figure}

The \textit{Field Line Universal relaXer} (FLUX) code provides a framework for modeling MHD evolution in the solar corona \citep{2007JASTP..69..116D}. `Fluxons’ are the piecewise-linear analogue for magnetic field lines, composed of segments connecting individual vertices. Each fluxon carries a finite quantity of magnetic flux. Within the simulation, forces are calculated and applied to vertex points, relaxing to an equilibrium state. Given the nature of the computed forces, the resulting output is a nonlinear force-free field. Figure~\ref{fig:fluxonmodel} illustrates a relaxed fluxon coronal model.

\subsubsection{What are fluxons?}

\begin{figure}[!htbp]
\plotone{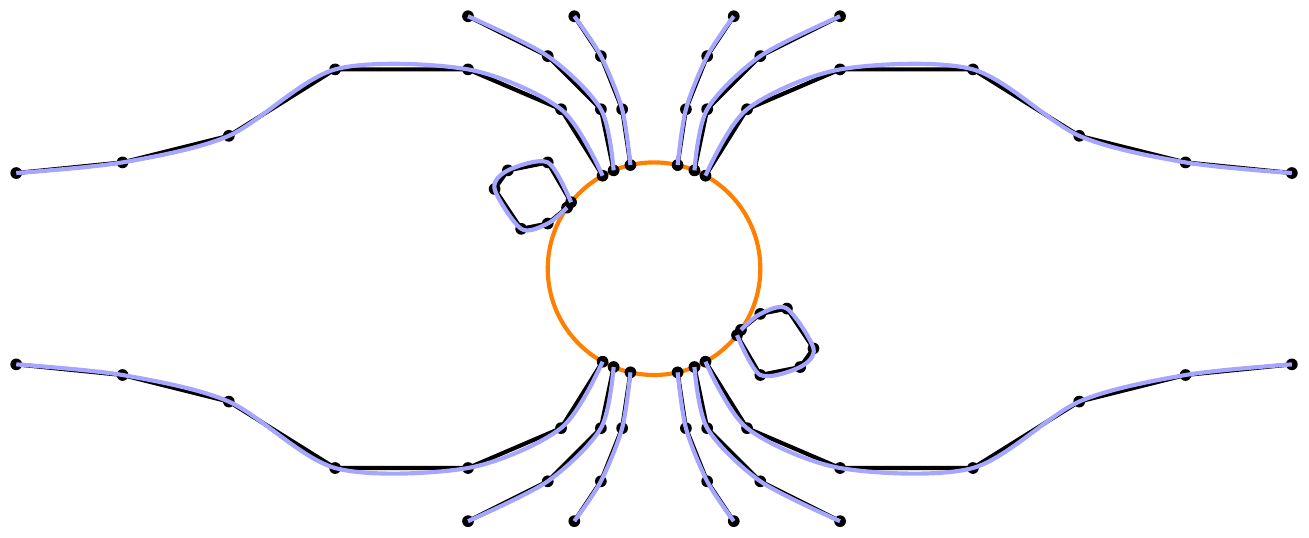}
\caption{\edit1{Schematic fluxon representation of solar magnetic field lines. Closed and open magnetic structures are displayed in blue with exaggerated scales.}}
\label{fig:fluxonschematic}
\end{figure}

Each fluxon structure represents a geometrical analogue for a bundle of magnetic fieldlines along a particular path through the solar corona. \edit1{Figure~\ref{fig:fluxonschematic} shows a cartoon schematic of a fluxon solar corona for reference.} A \edit1{definite} quantity of magnetic flux is contained along each fluxon. Each fluxon structure is broken down into a series of component parts, as outlined below \edit1{in a glossary of terms.}

\begin{itemize}
\item Fluxon - Solar magnetic fieldline analogue structure
\item Fluxel - Line segments along fluxons
\item Vertex - Points of intersection between fluxel segments
\end{itemize}

A fluxon structure itself remains intact through the relaxation process, unless manual reconnection is utilized to alter the topology of the system. However, fluxels and vertices along a given fluxon can be dynamically adjusted through relaxation, adapting to meet the geometrical requirements imposed. New fluxels and vertices are inserted to better approximate fieldline curvature without excessive angular changes between segments. In the \edit1{opposite} case, where fieldline curvature relaxes, excess fluxels and vertices are removed to reduce the computational burden.

With physical quantities only defined at points along each fluxon, interpolation is required for calculating magnetic and plasma parameters at arbitrary locations, or for interfacing with grid-based simulations. As with many interpolation schemes, the first step involves the selection of suitable points. For the simple case of falling directly atop a fluxon vertex point, that value can be directly assigned. In the more general case, for an arbitrary interpolant point within the simulation domain, ideally a 3-simplex of four points can be selected that enclose the spatial point in question. The interpolation algorithm searches the neighborhood of fluxon vertex points, and from nearby points constructs an ideal set of four vertex points that surround the interpolant point while minimizing the 3-simplex volume. The interpolated value is generated by weighing each vertex point by the volume of the 3-simplex formed by the remaining three neighboring points and the interpolant point itself.

For a sufficiently populated domain with sufficient fluxon vertex points from which to choose, creating an ideal neighborhood of points is generally not an issue. For unique geometries, or near the edge of fluxon occupied domain, suitable points can become more sparse. In these edge cases, the interpolation routine falls back in dimensionality, selecting and weighing from a 2-simplex, a 1-simplex, or in the rarest of cases a 0-simplex (from 3, 2, and 1 point, respectively). These cases run the risk of enhanced errors, and are toggled `off' by default.

\begin{figure}[!htbp]
\plottwo{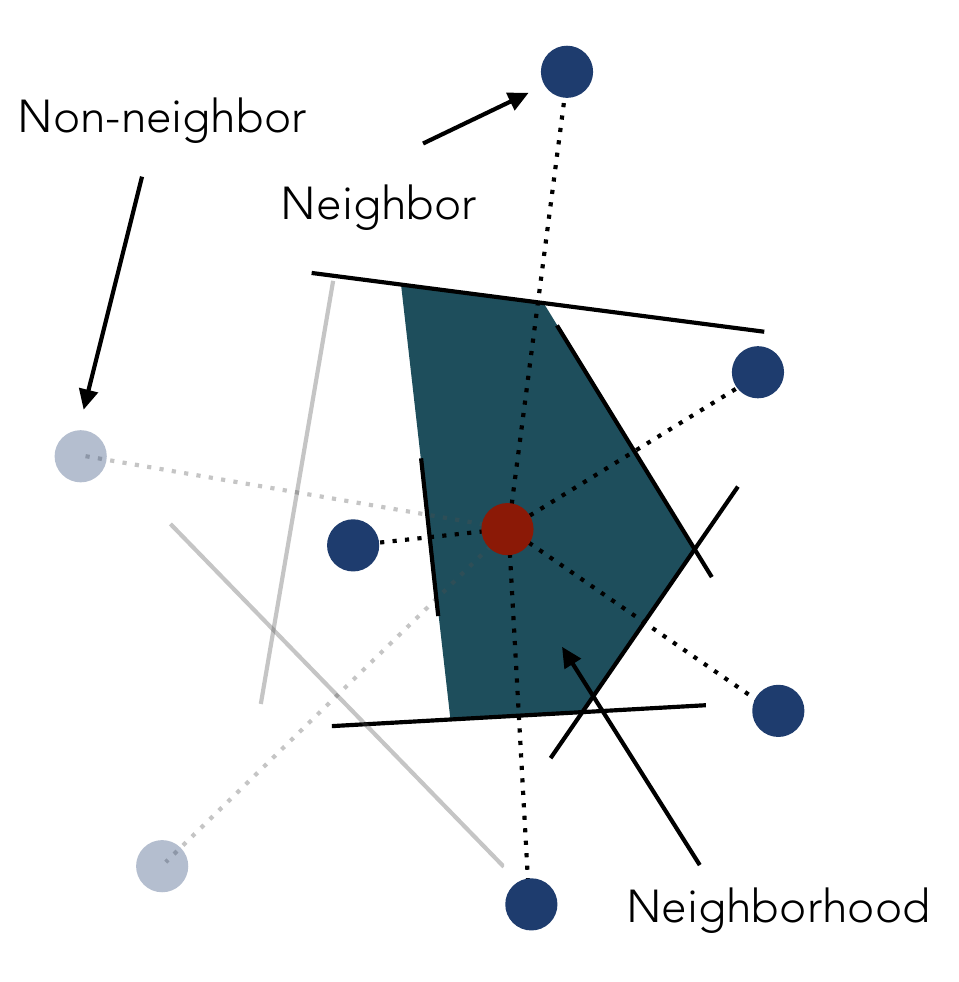}{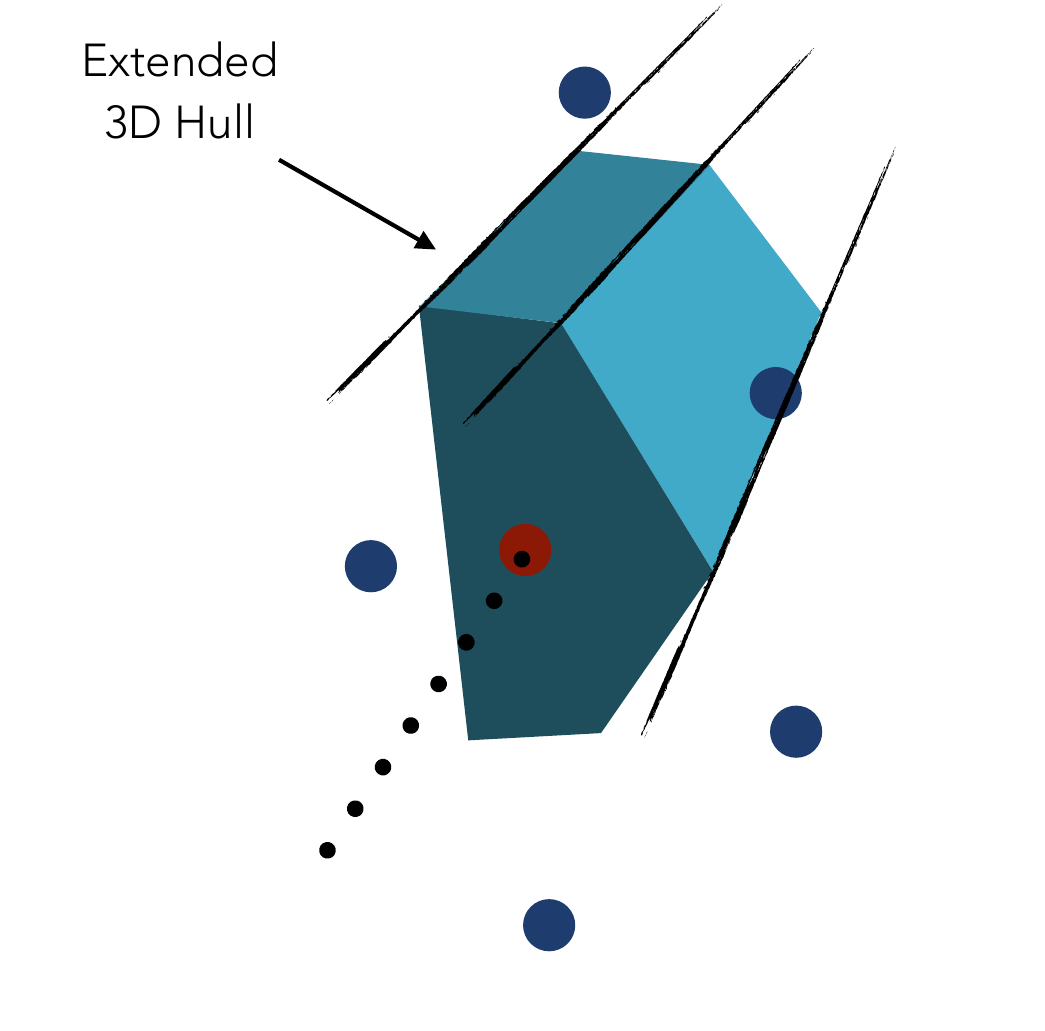}
\caption{Two-dimensional (left) and three-dimensional (right) representation of defining the neighborhood of a fluxel segment. A two-dimensional encapsulating polygon composed of perpendicular bisectors to neighboring fluxel projections is extended into a three-dimensional hull volume. This process is repeated for each fluxel segment along a given fluxon.}
\label{fig:neighborhood}
\end{figure}

\edit1{Though fluxons are one-dimensional objects, a volumetric representation of each fluxon is useful when computing the spatial domain of each element. Figure~\ref{fig:neighborhood} shows a schematic of this definition. For each fluxel element, a perpendicular plane is constructed, and intersections with nearby fluxons computed. From these intersecting points a minimum area two-dimensional polygon is defined from perpendicular bisectors, classifying adjacent fluxons as neighbors or non-neighbors. This two-dimensional polygon is extruded into three dimensions, defining a hull volume. This process is then repeated for each fluxel segment of a fluxon, and for every fluxon.}

\subsubsection{How do they relax?} \label{sec:howrelax}

The FLUX model works to relax a given initial magnetic topology at a given timestep to a force-free state. By calculating forces (magnetic pressure, magnetic tension from fieldline curvature, and a pseudo-force to space vertices along fluxons) at each vertex point from neighboring vertices and fluxel segments, these structures can be shifted in the net force direction, repeating this process until a sufficiently relaxed state is reached.

Equation~\ref{eqn:rlxcurv} describes the magnetic curvature force at each vertex point. $\Delta\theta_v$ \edit1{defines} the angle extended by the two fluxel segments extending from that vertex, and $l$ describes the length of the fluxel segment. Integrating along from the previous vertex to the next, the curvature force simply reduces to this angular difference between segments.

\begin{equation}
F_{cn,v} = \frac{1}{2}\int_{v-1}^{v+1}\mathcal{F}_{cn}~ds = \frac{l}{2}\frac{2\Delta\theta_v}{l}=\Delta \theta_v
\label{eqn:rlxcurv}
\end{equation}

\begin{figure}[!htbp]
\plotone{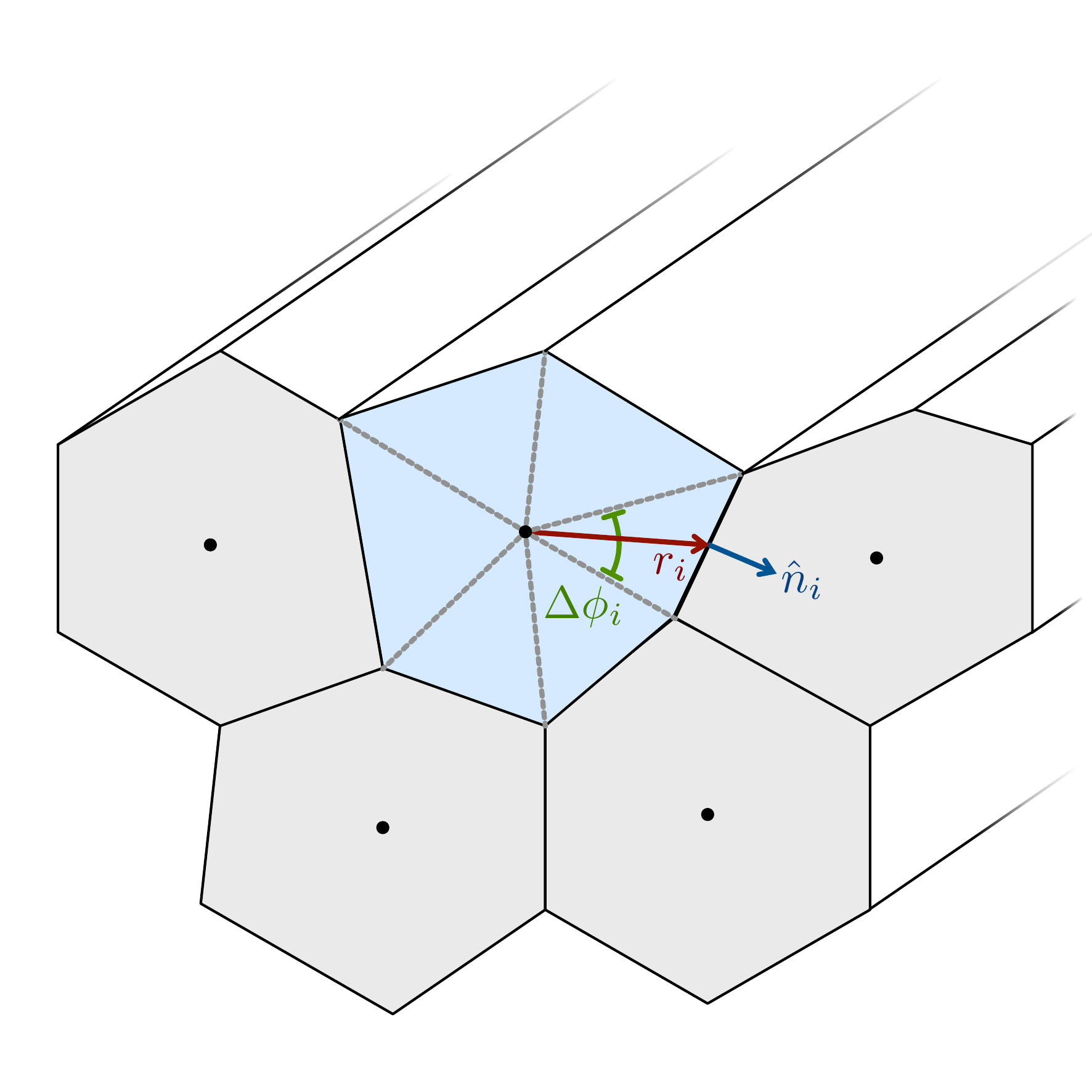}
\caption{Schematic diagram illustrating the cross-section of a fluxel hull segment, along with adjacent fluxel hulls. The angle \edit1{subtended} by each hull segment is given by $\Delta\phi_i$, with radius to the outer boundary as $r_i$, and the normal to the hull boundary given as $\hat{n_i}$.}
\label{fig:force_hull}
\end{figure}

Equation~\ref{eqn:rlxpres} describes the magnetic pressure force exerted on each fluxel segment. The integration is performed from one vertex point to the next, along the length of the fluxel, and reduces down to a perpindicular gradient of the magnetic field. Due to the nature of the fluxon model, the magnetic field within each hull cross-section of a fluxel is simply the magnetic flux subdivided into angular wedges made from tesselation with neighboring fluxels. This term then simply reduces to a sum over each hull segment $i$, involving the normal to each wedge segment boundary $\hat{n_i}$, the angular extent of each $\Delta\phi_i$, and the radius out to that hull boundary edge $r_i$. \edit1{Figure~\ref{fig:force_hull} displays these geometric elements in the context of a fluxel hull segment.}

\begin{equation}
\mathbf{F}_{pn} = \int_{v}^{v+1}\mathcal{F}_{pn}~ds = l \frac{\langle \nabla_\perp B \rangle}{\langle B \rangle} = l \sum_i(\hat{n}_i\Delta\phi_i)/(\pi r_i)
\label{eqn:rlxpres}
\end{equation}

\begin{figure}[!htbp]
\plotone{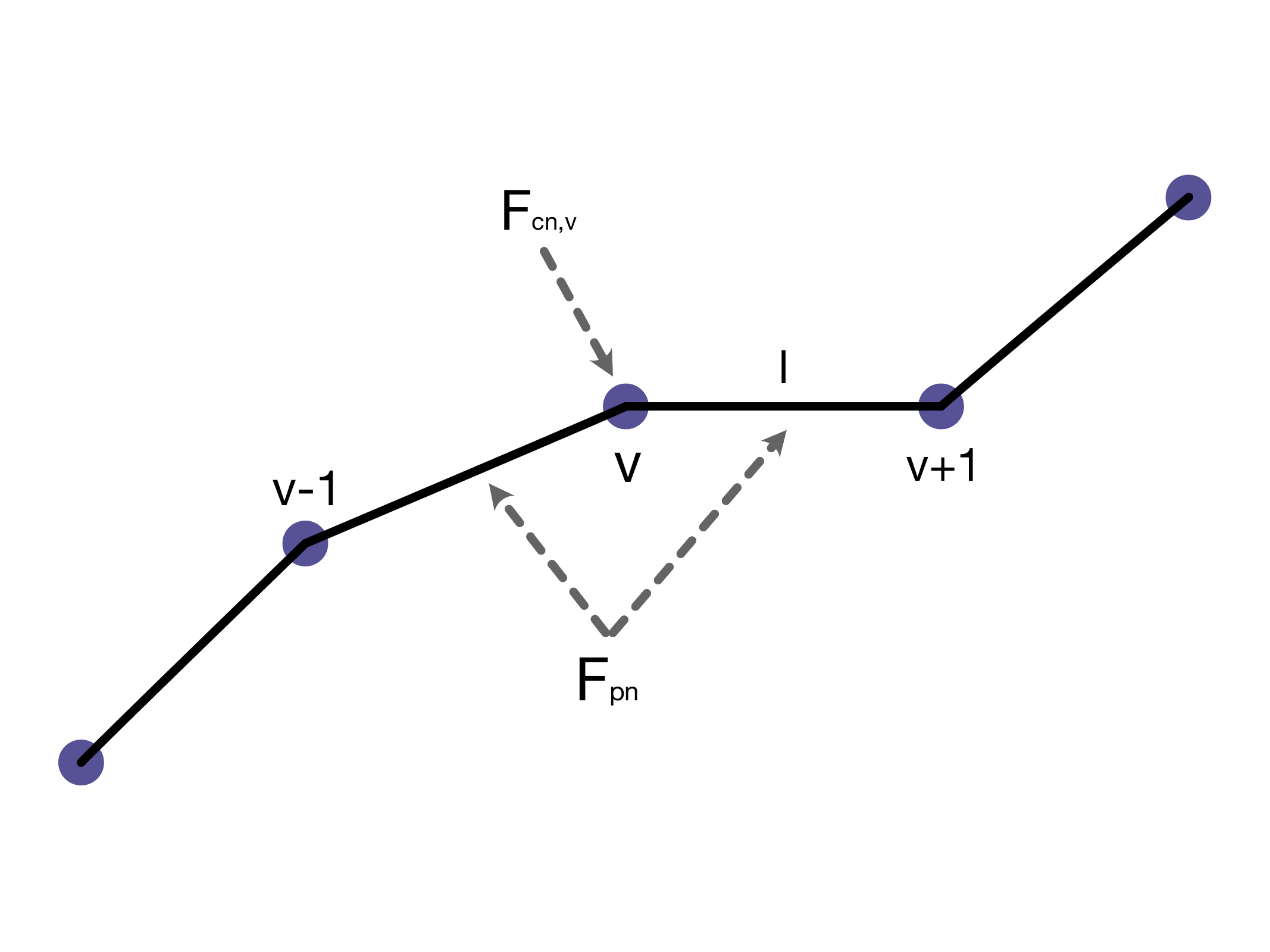}
\caption{Schematic diagram illustrating a given vertex point $v$, and adjacent fluxel segments. The acting locations for both the magnetic curvature force, $F_{cn,v}$, and the magnetic pressure force, $F_{pn}$, are displayed.}
\label{fig:fluxon_forces}
\end{figure}

Considering these two forces described, the net force on each vertex point is given by Equation~\ref{eqn:rlxverfor}. The curvature force applies directly to the vertex point. The second set of terms describes the average force between the two fluxel segments neighboring the vertex point in question. \edit1{Figure~\ref{fig:fluxon_forces} shows a schematic of how these forces apply for a given vertex point.}

\begin{equation}
\mathbf{F}_v = \mathbf{F}_{cn,v} + \frac{1}{2} \left(\mathbf{F}_{pn,f} + \mathbf{F}_{pn,f-1}\right)
\label{eqn:rlxverfor}
\end{equation}

Finally, the computed force at each vertex point is used to shift said vertex point in space, moving towards relaxation. Equation~\ref{eqn:rlxvershf} describes this shift, as a quasi-Eulerian step. Each of the forces $F_j$ are summed over. The squared term in parentheses acts as a stiffness coefficient, preventing endless corrections near equilibrium. \edit1{The remaining} terms describe the step in space $\partial \tau$ and the distance to the nearest neighbor \edit1{$r_{min,i}$} (within neighboring fluxels). Together, these terms work to relax to a near force-free state, while controlling stiffness to prevent oscillation around equilibrium.

\begin{equation}
\Delta \mathbf{x}_i = \delta \tau \left( \frac{|\sum \mathbf{F}_j|}{\sum|\mathbf{F}_j|} \right)^2 r_{min,i} \sum \mathbf{F}_j
\label{eqn:rlxvershf}
\end{equation}

During this process the structure of each fluxon can dynamically adjust to the given geometry. Where the curvature between fluxel segments exceeds a user-defined threshold, additional vertex points and fluxels are dynamically inserted to allow for this curvature to relax. Additionally, where a given curvature is over-defined by accumulated vertex points and fluxels, these are consolidated to reduce computational burden during the relaxation process. \edit1{A further description of the relaxation process is found in the original methodology paper \citep{2007JASTP..69..116D}.}

\subsubsection{How can we model a data-driven corona?}

The FLUX model allows for placement of fluxon structures extending from an inner photospheric boundary at 1~R$_\odot$ out to 21.5~R$_\odot$ (0.1~AU). Note that while open fluxon structures are rooted to the photosphere at 1~R$_\odot$ on one end, the remaining end is free to relax to an equilibrium position. \edit1{This outer `boundary' at 21.5~R$_\odot$ is used to provide a surface on which to interpolate values of intersecting fluxons to compare with other models and observations, and to demarcate open fluxon structures. This surface does not restrict fluxon movement or positioning, as fieldlines are free to relax while intersecting this surface.}

Input magnetograms can be utilized to setup an initial fluxon `world' for relaxation. Here the term world refers to the data structure containing appropriate boundary conditions, fluxons (and contained vertex points), along with relational data. This world structure is the analogue in the fluxon model of the grid-based simulation domain. This world is allowed to relax and respond to internally generated forces.

Taking an input magnetogram, the distribution of magnetic flux can be segmented into discrete fluxons using a form of dithering. Dithering, which has etymological roots in manual vibrations reducing error in mechanical computers, can be used in a digital context for purposes of converting a \edit1{grayscale} image to a binary black-and-white version. The error created in each pixel conversion is shifted along a defined pattern, distributing the error across the image. This method is suited for the conversion of magnetogram distributions of magnetic flux (ranging scalar values in each pixel) into a map of fluxon locations (each fluxon carrying discrete units of magnetic flux). To avoid a directional bias when dithering input magnetograms, the area of each map is read using a Hilbert curve tracing \citep{hilbert1891}. \edit1{For more details on using Hilbert curves for initializing fluxon placement see \cite{2010ApJ...715.1556R}.}

The FLUX model is able to assimilate various types of magnetic input data, including synoptic magnetograms, synchronic magnetograms, flux transport model data, or combinations thereof. These initial states are then allowed to relax given appropriate internal magnetic forces from fluxon structures. The resulting time series represents a series of nearly force-free equilibria, wherein fluxons structures can be linked and tracked.

The combination of an input magnetogram and seeded fluxon footprint locations can be used to define an initial three-dimensional fluxon distribution. These two data sets are used to trace the corresponding fluxons into the coronal volume using \edit1{a potential field source surface (PFSS) model - \texttt{pfsspy}} \citep{Stansby2020}. This sets the initial topological state of the model, which then undergoes subsequent relaxation to a nonlinear force-free field. \edit1{Note that although the initial open / closed topology is set by a PFSS model, the resulting relaxed field solution is \edit1{a Nonlinear Force Free Field (NLFFF)} in nature, and extends far beyond the typical source surface of a PFSS model. The FLUX model contains the ability to manually reconnect field line structures, which is of particular interest to study the nature of open / closed boundary reconnection in future work.}

One of the direct benefits of this assimilation methodology is the detailed mapping of open magnetic flux linkages. Of particular interest is the detailed response of the quantity and distribution of open magnetic flux to changes in the underlying magnetic configuration (emergence, cancellation, migration, etc). With direct mapping along each fluxon structure from source to open end, these relations can be studied in more robust detail.

\subsection{Solar wind modeling} \label{sec:windmodeling}

A one-dimensional open fluxon provides a unique structure on which to compute a solar wind profile. The hull area profile of each fluxon is stored along the length of it, providing an expansion factor.

\begin{figure}[!htbp]
\epsscale{1.2}\plotone{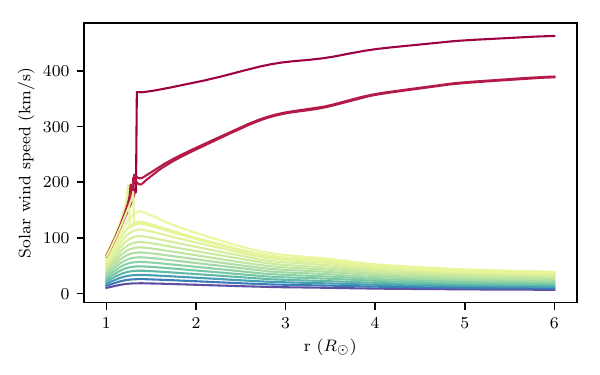}
\caption{Example of binary search tree method for finding a transonic solar wind velocity profile for a given open fluxon. An initial solar breeze solution is found (blue), with increasingly warm colors honing in on the ultimate stable transonic solution (red).}
\label{fig:wprof}
\end{figure}

If we start with a modified Parker 1D solar wind solution,

\begin{equation}
\frac{d v}{d r} = \frac{v}{r} ~ \frac{r g - 2 c_s^2 \left[ 1 + \frac{r}{2 f_r} \left( \frac{df_r}{dr} \right) \right]}{c_s^2 - v^2},
\label{eqn:solarwind1d}
\end{equation}
where our sound speed $c_s$ and expansion factor $f(r)$ are defined by

\begin{equation}
c_s = \sqrt{\frac{2kT}{\mu m_p}} \quad \text{and} \quad f(r) = \frac{A(r)}{A_0} \left( \frac{r_0^2}{r^2} \right),
\label{eqn:solarwind1d_def}
\end{equation}

we can iterate to find a transonic solution for each open fluxon given an initial set of boundary conditions at the root. \edit1{Each fluxon has an initial base temperature of 1e6~K.} A binary search tree is utilized to seek out this solution for each open fluxon by adjusting these initial parameters at the base, with an example of such a search illustrated in Figure~\ref{fig:wprof}. \edit1{Note in particular the sequence of solar `breeze' solutions - those indicated from blue to yellow. These solutions fail to reach the critical supersonic point. Also note the large discontinuity plotted in red - a nonphysical solution encountered through the numerical integration process. A singularity in Equation~\ref{eqn:solarwind1d} at the sound speed requires the use of an adaptive integration technique to bypass this point. After iterating with this search tree and discarding invalid solutions, the algorithm hones in towards a final velocity profile.} Given a series of open fluxon structures, the methodology outlined above can be repeated en masse, mapping an ensemble of transonic solar wind solutions onto a two-dimensional output surface.

\section{FLUXPipe} \label{sec:fluxpipe}

FLUXPipe represents a pipelinification of the FLUX code, such that one can specify a Carrington Rotation of interest and a number of magnetic footpoints, and the code will \edit1{carry} out the entire pipeline with no additional input.

As an overview, the pipeline performs these steps, which will be described in the following subsections:
\begin{enumerate}
\item Retrieve and process a polar corrected HMI synoptic chart of radial magnetic field strength.
\item Trace a Hilbert curve across the image to find footpoints that represent equal amounts of magnetic flux.
\item Trace the footpoints into the corona using the \texttt{pfsspy.pfss} function, determining an initial magnetic topology.
\item Use the FLUX relaxation code to relax the fields into a realistic topology and near force-free state, but allowing for currents (such that it is no longer a potential field).
\item Generate solutions for the solar wind velocity within the open magnetic field lines.
\end{enumerate}

This effectively allows \edit1{one} to use the fluxon code in a straightforward way to output a data-driven solar wind map generated at a 21.5~R$_\odot$ radius surface.

\begin{figure}[!htbp]
\epsscale{1.2}\plotone{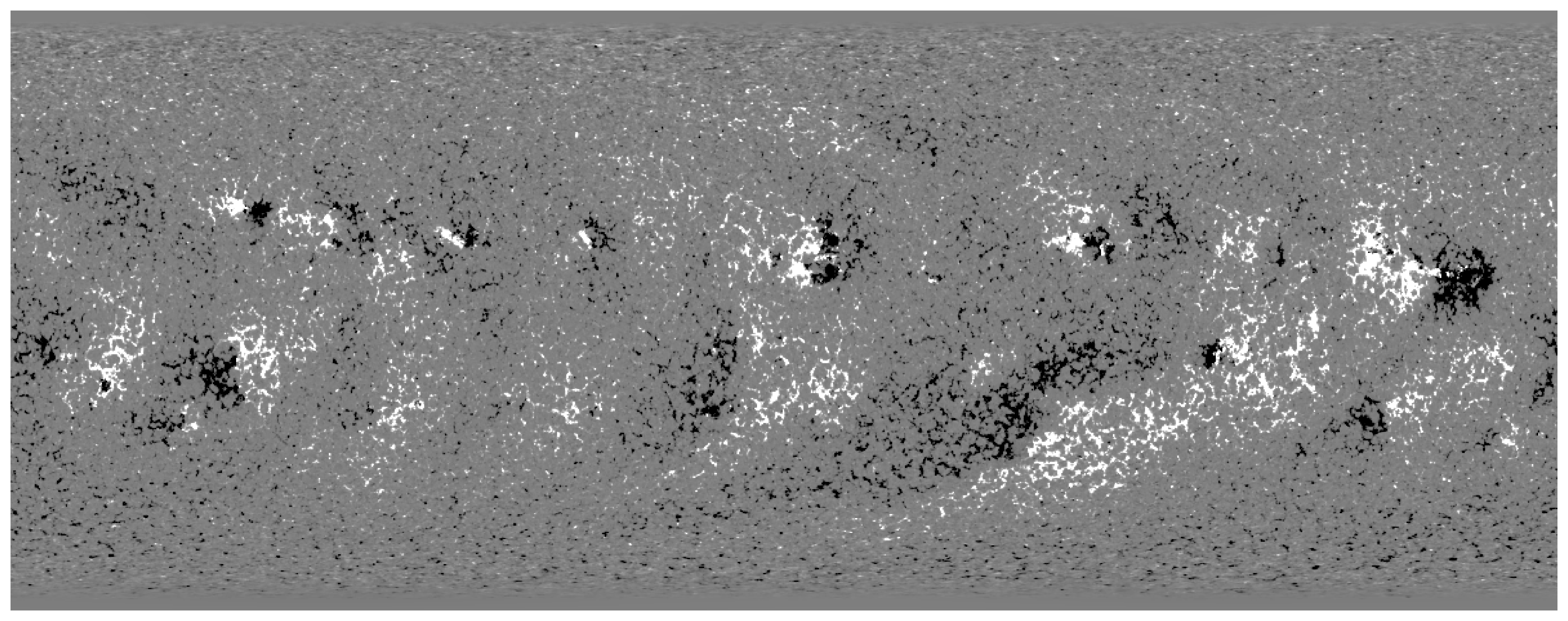}\\
\epsscale{1.2}\plotone{fig_magnetogram_b.png}
\caption{\textit{(top)} Input HMI polar-corrected synoptic magnetogram for Carrington Rotation 2160. \textit{(bottom)} Footpoint placement of fluxons using the Hilbert curve methodology. Open footpoints are mapped in red (positive) and blue (negative), with closed footpoints mapped in orange (positive) and teal (negative).}
\label{fig:magnetogram}
\end{figure}

\subsection{Pipeline usage}
To run the code, one simply calls the \edit1{\texttt{config\_runner.py}} file with the Carrington Rotation number of interest as input. One can specify a single rotation or a list of rotations\edit1{, as well as several other parameters, in the \texttt{config.ini} file}. The runner file then calls \texttt{magnetogram2wind.pdl}, which is the main pipeline script, and utilizes both PDL and Python libraries \edit1{and scripts}. In the following subsections we will walk through each step of the FLUXPipe algorithm using Carrington Rotation 2160 as an example case.

\subsubsection{Data retrieval and processing}
An input \edit1{polar-field-corrected} HMI synoptic chart of radial magnetic field strength is retrieved using the Joint Science Operations Center (JSOC) through SunPy. It is then downsampled using the \texttt{astropy.nddata.block\_reduce} function with \edit1{\texttt{numpy.nansum}} as the reduction function. \edit1{This map is} plotted in Figure~\ref{fig:magnetogram} for reference.

\edit1{\subsubsection{Fluxon footpoints}}
The magnetic map is then traced with a Hilbert curve to find footpoints that represent equal amounts of magnetic flux. The equal-flux value is iterated using Eulerian minimization until the number of defined footpoints is close to the number requested by the user. \edit1{The second panel of} Figure \ref{fig:magnetogram} shows these footpoints for CR2160.

\begin{figure}[]
\includegraphics[width=0.48\textwidth,trim=3cm 2cm 3cm 5cm, clip]{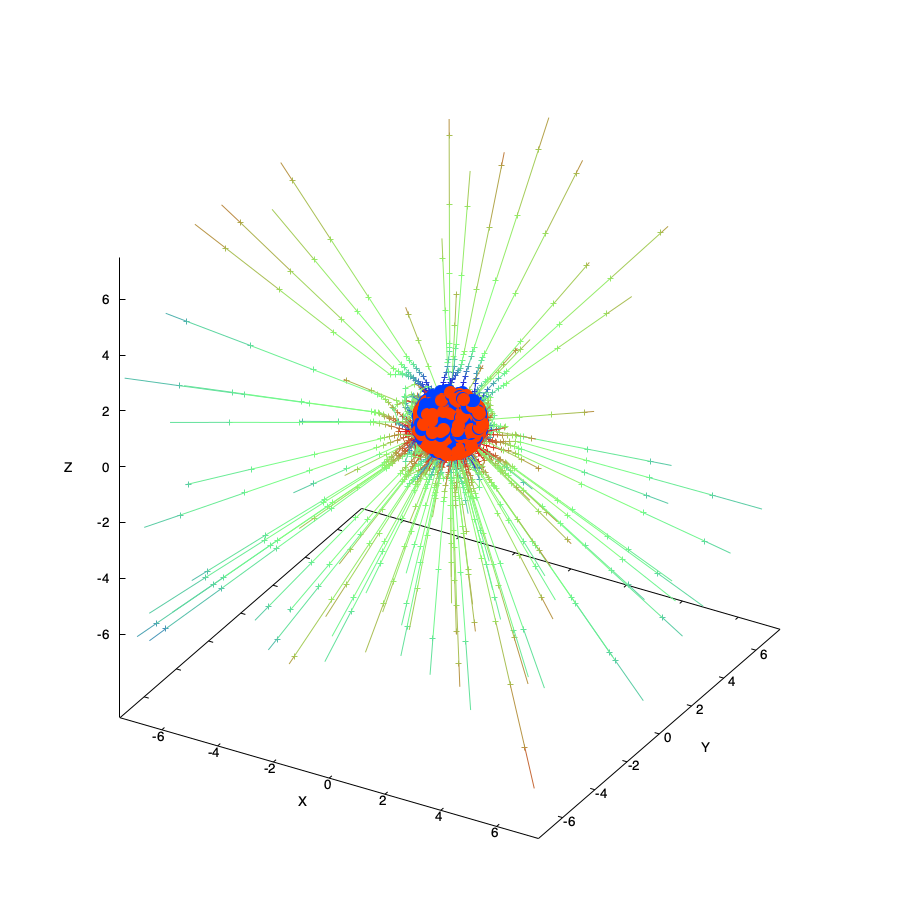}\\
\includegraphics[width=0.48\textwidth,trim=3cm 2cm 3cm 5cm, clip]{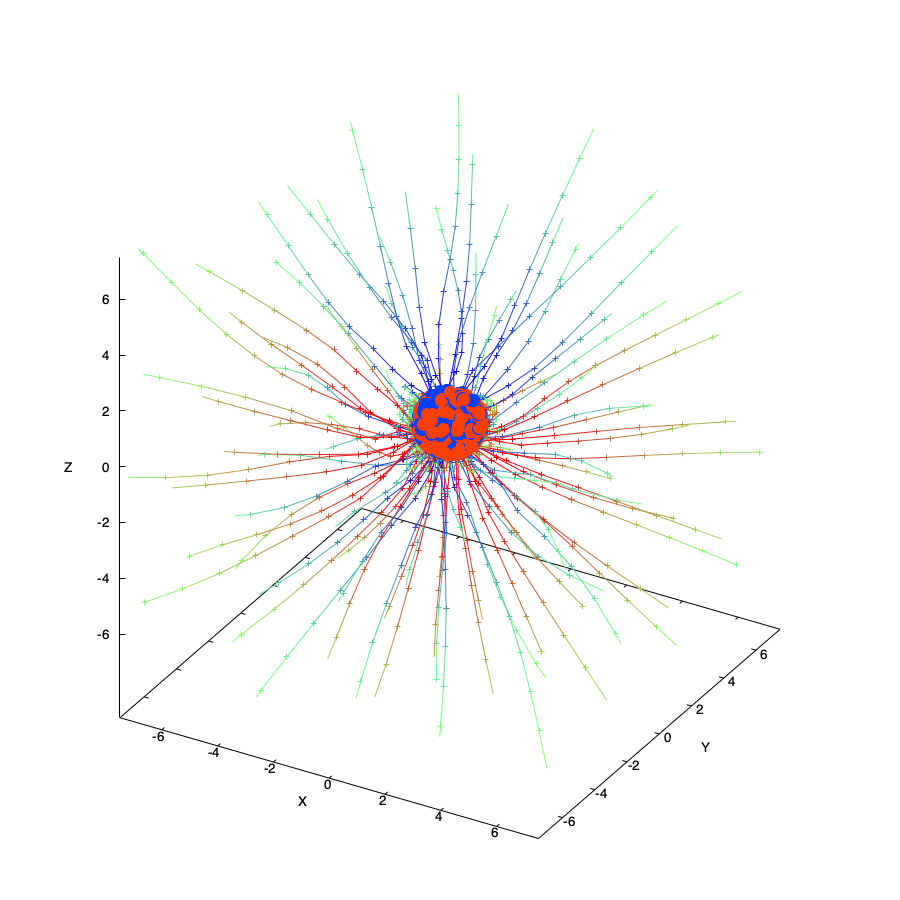}
\caption{Carrington rotation 2160: \textit{(upper)} The fluxon simulation as initialized. \textit{(bottom)} The fluxon simulation after 2000 steps of relaxation, in a state of transition towards a fully relaxed force-free field state.}
\label{fig:initial_relaxed}
\end{figure}
    
\subsubsection{Topology initialization}
The footpoints that result from the Hilbert algorithm are traced into the corona using the \texttt{pfsspy.pfss} function. A tracing function is defined from the input magnetogram, after which the initialized footpoints are traced into the corona, and are separated into open and closed fieldlines. Beyond 2.5~R$_\odot$, open fieldlines are extended radially out to 21.5~R$_\odot$. This provides an initial condition and input topology \edit1{of a PFSS} magnetic field geometry, which is then ingested by the FLUX code to generate a collection of fluxons that can be used by FLUX. \edit1{The first panel in} Figure \ref{fig:initial_relaxed} shows this initial unrelaxed world.

\subsubsection{Relaxation}
As outlined in \S\ref{sec:howrelax}, forces are computed along each fluxon in the simulation, with spatial coordinates shifted in relaxation space towards a force-free equilibrium. This relaxes the fields into a realistic topology and near force-free state, but allows for non-potential fields by allowing currents. \edit1{The second panel of} Figure \ref{fig:initial_relaxed} shows the result of 2000 steps of relaxation.

\subsubsection{Wind calculation}
Open fluxons are gathered \edit1{and} the solar wind calculation begins as outlined in \S\ref{sec:windmodeling}. For each open fluxon, the one-dimensional Parker solar wind equation \edit1{(Equation \ref{eqn:solarwind1d})} is integrated along its length, using the computed expansion factor along the way. A binary search tree integration method is used to hone in on a transonic solution. The resulting solar wind solutions are gathered together as an ensemble, and mapped at their intersection with a spherical surface at 21.5~R$_\odot$. \edit1{Other methods for calculating the solar wind are also being implemented for model verification and to meet future needs.}

\begin{figure*}[!ht]
\includegraphics[width=0.49\textwidth]{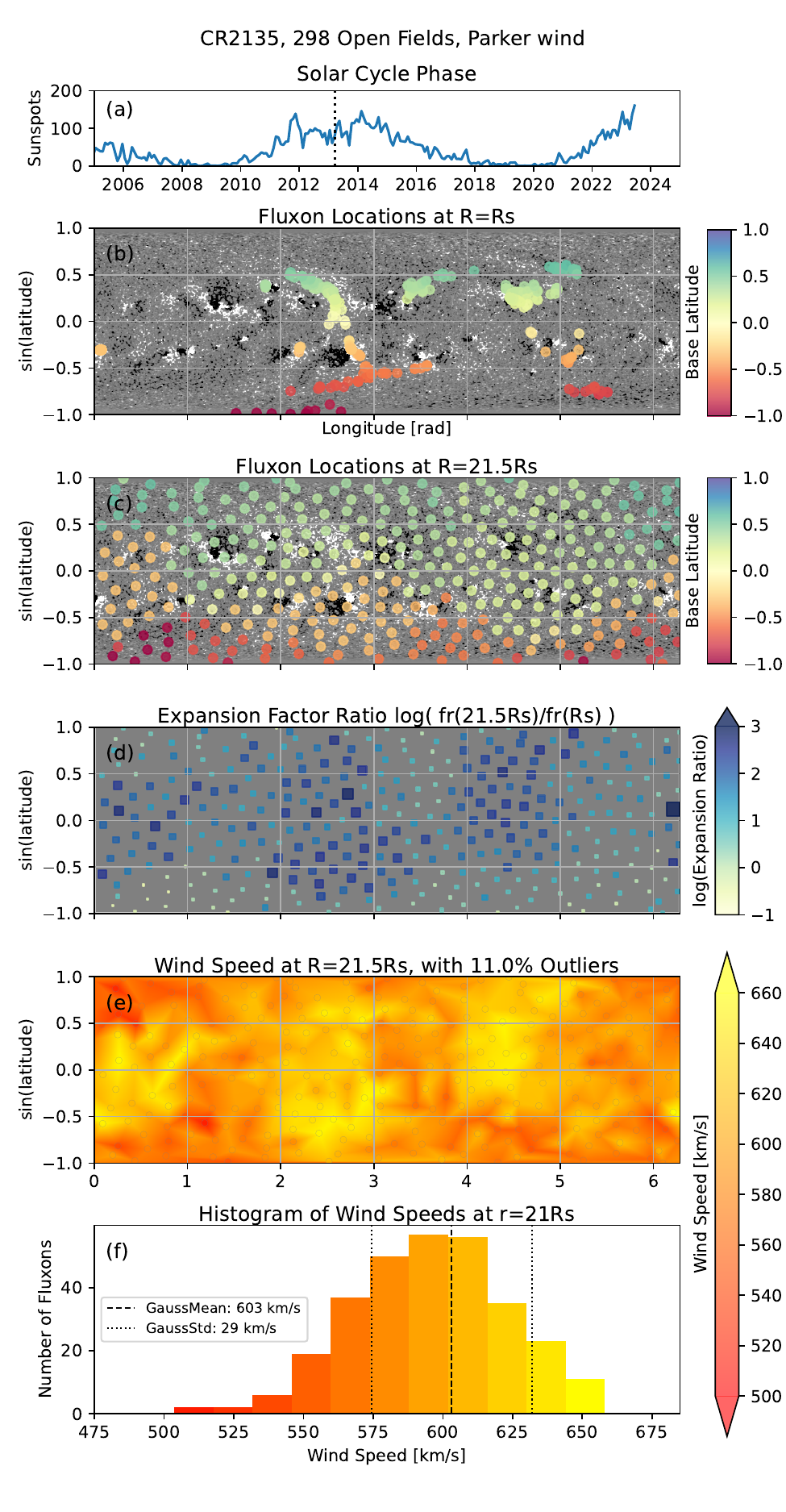}
\includegraphics[width=0.49\textwidth]{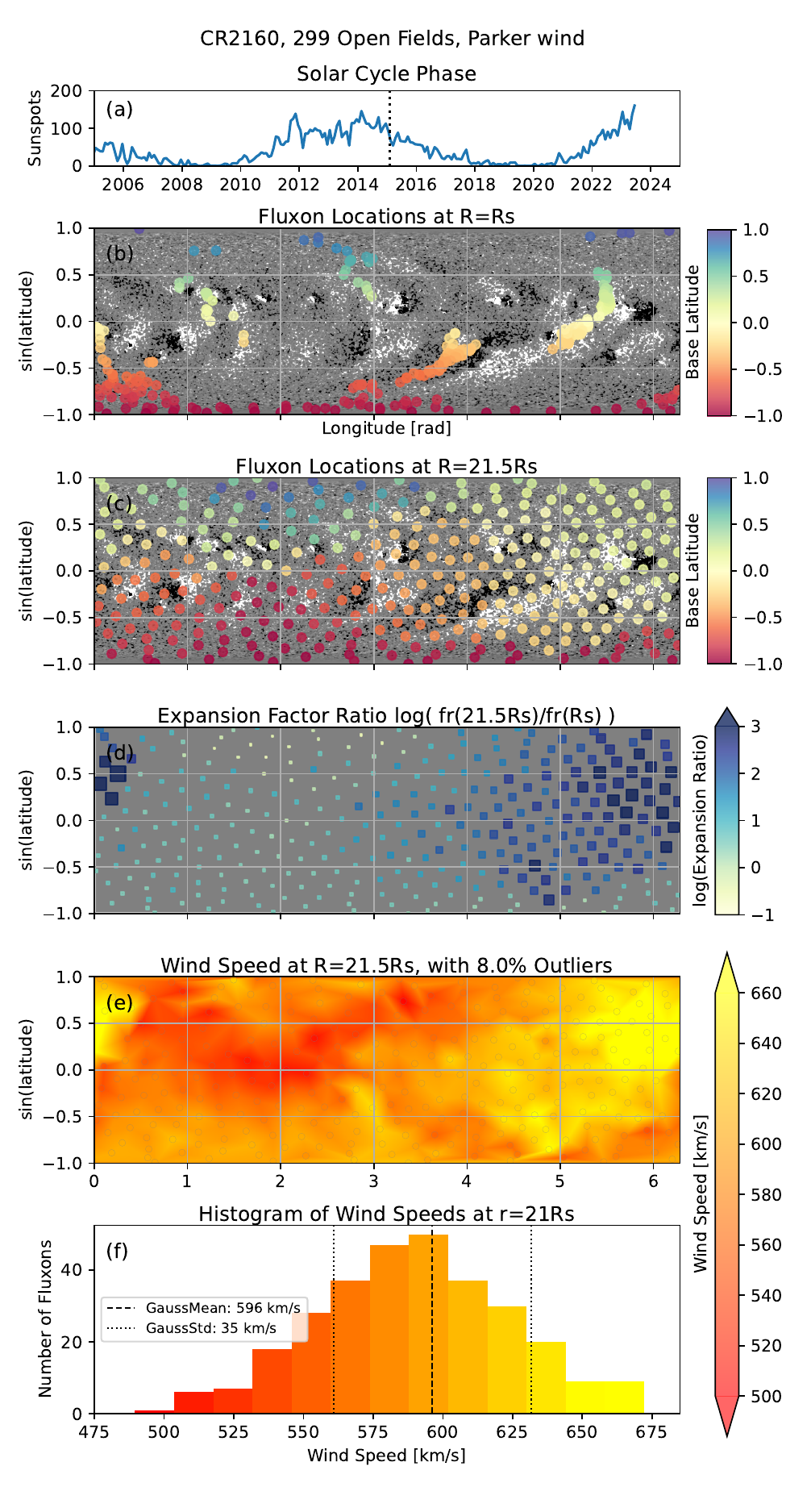}\\
\caption{
\edit1{Results of running FLUXPipe on Carrington rotations CR2135 and CR2160. These rotations were at the middle and ending of solar maximum, respectively. 
(a) The location within the solar cycle is marked with a vertical dotted line. 
(b) Footpoints of open fluxons overlying the HMI magnetogram, with color representing the footpoint latitude. 
(c) The location of the fluxons on the outer boundary at $R = 21.5 R_s$. The colors match the latitude of the fluxon footpoints, allowing for correlations to be drawn. 
(d) The log of the ratio of the expansion factor (Equation \ref{eqn:solarwind1d_def}) between the two altitudes. 
(e) the solar wind speed as determined by integrating Equation \ref{eqn:solarwind1d}. 
(f) A histogram of those wind speeds.}}
\label{fig:tri-results}
\end{figure*}

\begin{figure*}[!ht]
\includegraphics[width=0.49\textwidth]{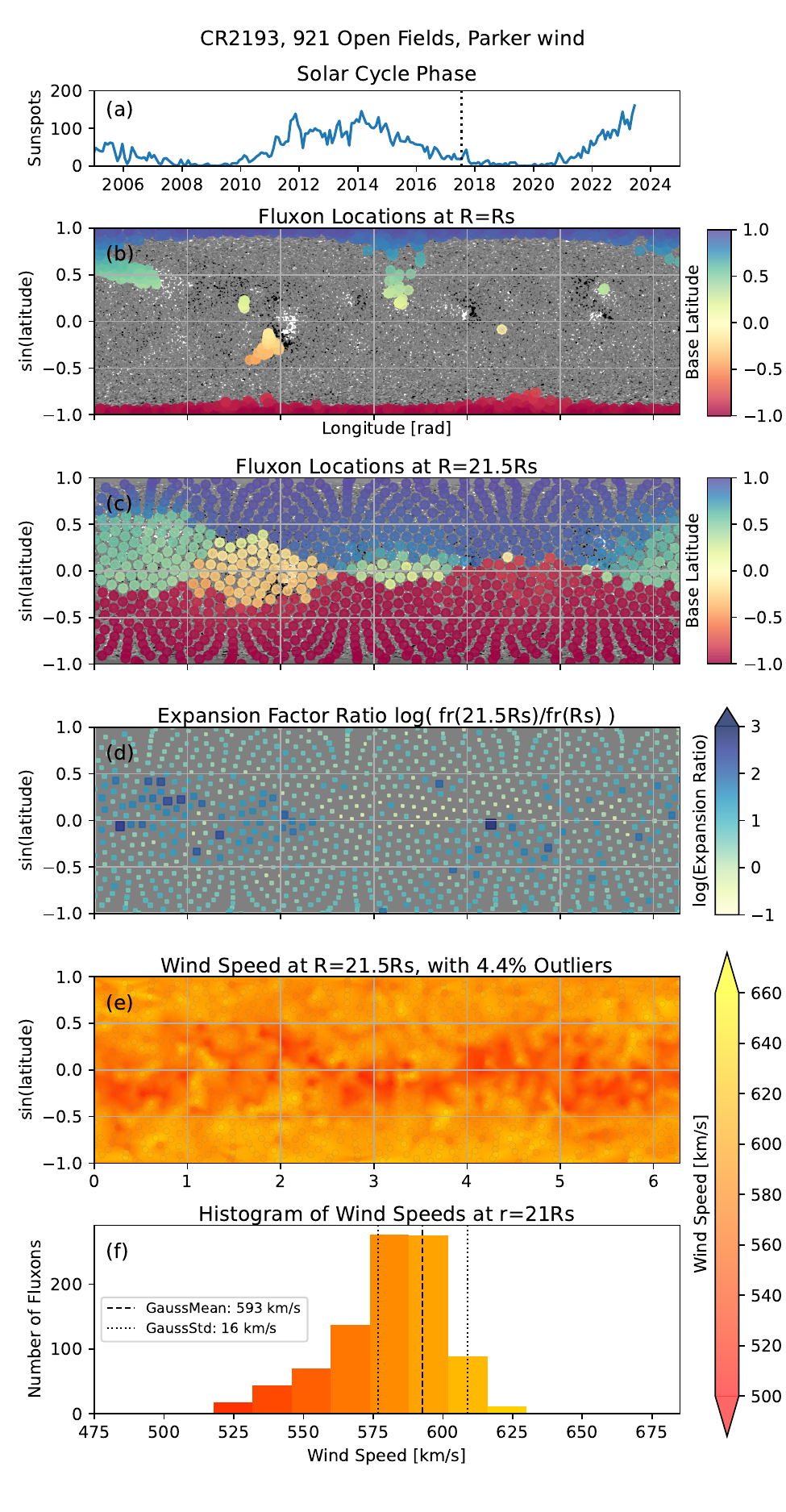}
\includegraphics[width=0.49\textwidth]{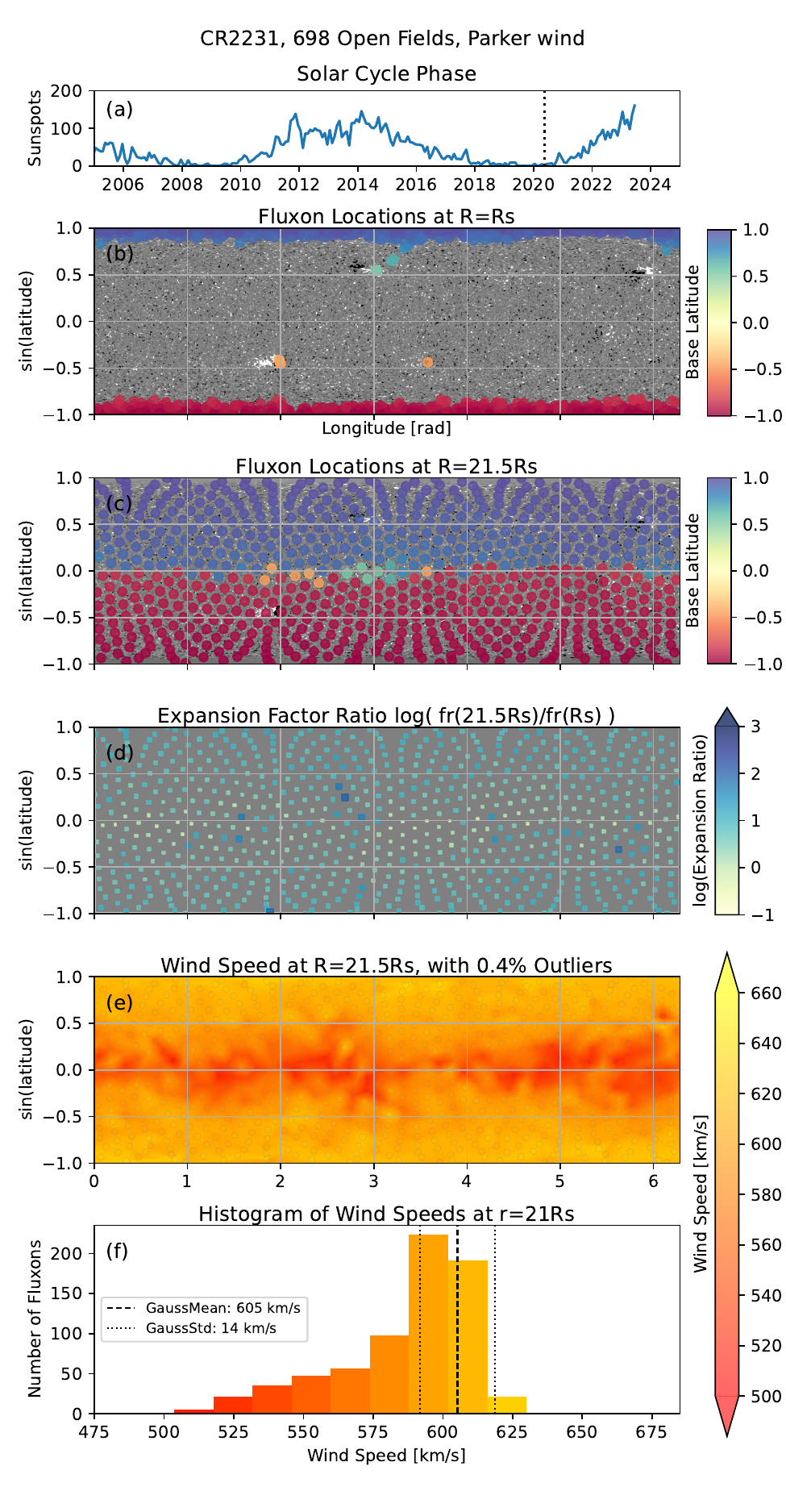}\\
\caption{\label{fig:tri-results-2} Results of FLUXPipe on Carrington rotations CR2193 and CR2231, before and after solar minimum. Panels are as described in Figure \ref{fig:tri-results}. As indicated in the top panel, CR2193 sits at the beginning of solar minimum, and CR2231 lies near the end of solar minimum.}
\end{figure*}

\section{Results} \label{sec:results}
To demonstrate the \edit1{output} of the FLUX code we have chosen \edit1{to examine} four Carrington rotations of interest. These rotations mark specific points in the rise and fall of the solar magnetic activity cycle. \edit1{The behavior of the solutions fall into two relatively distinct groups as shown in Figure~\ref{fig:tri-results} and Figure~\ref{fig:tri-results-2}.}

\edit1{In these figures, the panels are as follows: Panel (a) shows what at which point in the solar cycle the Carrington rotation lies, allowing for a general understanding of the context of each figure. Panel (b) shows the locations of the footpoints of the open magnetic fields at the base of the simulation. The color of these points corresponds to the latitude of the footpoints, which becomes useful in the following panel to understand the source region of the various features in the plot. Panel (c) shows the locations of the fluxons at the outer boundary of the simulation. Because the color of each point is the same as in panel (b), one can correlate the lower and upper points of specific bundles of fluxons. Panel (d) displays the log ratio of the expansion factors at the bottom and top of the domain. Darker blue points indicate more expansion. The size of the points also grows as a function of the expansion ratio. In essence, this panel shows how much expansion the fluxons have undergone in their path from the lowest to the highest part of the domain. Panel e) shows the calculated wind speed at the top of the domain, which is strongly correlated with the expansion factor. For panels (b-e) we use a sine latitude projection to maintain equal area per pixel for comparison with other maps, and to not overemphasise the poles. Panel (f) presents a histogram of those wind speeds.}

\edit1{\subsection{Model observations}}
\edit1{Between all of these rotations, the distribution of solar wind speeds maintains a clustering around \edit1{600}~km/s, with faster and slower outliers as a result of the nozzling effect of the changing computed expansion factor along each open fluxon.}

Carrington rotation \edit1{2135}, \edit1{occurring} towards the \edit1{peak} of solar activity cycle 24, displays relatively little open polar magnetic field; \edit1{The open field lines are all rooted in equatorial regions. The various features in the equatorial zone expand to fill the space, but it isn't possible to predict the wind speed by considering footpoint location alone. Panels d) and e) demonstrate the correlation between the solar wind speed and the expansion factor ratio.}

Carrington rotation 2160, on the other side of the peak of solar activity cycle 24, displays somewhat similar behavior. In this case, however, a more distinct patch of open magnetic field is rooted at the southern pole. \edit1{These polar fields take up more of the outer boundary, leading to moderate wind speeds in that region. On the right side of the plot, however, a group of open field lines from the equator have expanded by several orders of magnitude, and this has led to a large region of much faster wind speeds. Again, the expansion factor plays a key role in determining the acceleration of solar wind outflows.}

Carrington rotation 2193, situated in the declining phase of solar cycle 24, shows remarkably differing behavior. The majority of open fluxons are rooted at both the north and south poles, with pockets of open field at lower latitudes. \edit1{In this case, the polar fields dominate the domain, and the equator to pole variation of wind speed is the dominant feature.} \edit1{We do see the expansion effects of small parcels of equatorial open magnetic field, particularly the isolated patch near 2~radians (115~degrees). As this expands it leave a significant trace in the map of expansion factors.}

Carrington rotation 2231, placed right as solar cycle 25 begins to ramp up, shows similar behavior to \edit1{CR 2193.} Most of the open fluxons are rooted towards the poles, with in this case fewer open fluxons towards the equatorial regions. \edit1{Here we see the effects of some isolated patches of equatorial open field, but without large scale effects in the map of expansion factor. In both rotations 2193 and 2231, the dominant behavior of the outflow wind maps is a peak near each of the polar regions, with a progression towards a relatively slower meandering band of outflow wind speeds near the equator.}

\section{Discussion} \label{sec:discussion}

Fluxon modeling allows a unique approach to simulating the solar coronal magnetic field, yielding a number of advantages. The discretized nature of fluxon magnetic fieldlines allows for rapid assimilation, relaxation, and solar wind modeling. It avoids many of the grid-based issues that arise from numerical grid-based issues, such as numerical diffusion. Fluxons preserve the magnetic topology, unless magnetic reconnection is specifically triggered, preserving fieldline quantities such as winding.

The one-dimensional nature of fluxon magnetic fieldlines also provides the ideal case for analyzing the resulting solar wind acceleration. From the relaxed coronal magnetic configuration, each topologically open fluxon is a one-dimensional analogue of a bundle of open magnetic fieldlines. The spatial geometry, location of nearby neighbors, and magnetic expansion factor can be used to iteratively solve for a transonic solar wind solution.

Significant enhancements have been made towards leveraging FLUX as a pipeline tool for analyzing the solar wind, including the development of FLUXPipe for automated processing. The underlying code has been multithreaded, to allow for relaxation and solar wind computation more quickly.

\subsection{Code and data availability}

The FLUX code base is maintained as an open source GitHub repository, preserving the commit history of the code throughout its development lifetime. This contains installation instructions and a user-guide wiki set of documentation \citep{fluxonmhd}. The codebase itself is a mixture of underlying C code doing the heavy computational lifting, and overarching Perl and Python code for interfacing with the simulation.

Given the unique nature of the fluxon modeling approach, sharing data is not always a direct process. Data export is possible to a regular uniform three-dimensional grid for sharing / comparisons with other model data. Solar wind data is also interpolated onto a two-dimensional spherical surface for interfacing with other wind models.

\section{Conclusions} \label{sec:conclusions}

The FLUX model provides an intermediate approach between rapid heuristic methods and intensive 3D magnetohydrodynamic models, providing the best of both worlds. The FLUX model has the distinct advantages of being computationally efficient (scaling with the magnetic complexity of the two-dimensional photospheric boundary) and preserving connectivity to allow for tracking the history of a bundle of magnetic flux. Open fluxons extending from the photospheric boundary are used to compute a set of modified one-dimensional isothermal Parker solar wind solutions, with transonic solutions interpolated to an outer spherical boundary uniform grid at 21.5~R$_\odot$.

The FLUXPipe routine expands on this by providing an automated methodology for running a FLUX solar wind mapping from starting at input magnetogram assimilation to ending with output wind map gridding. This methodology was used to analyze four solar rotations throughout the solar cycle to catalog solar wind behavior.

\subsection{Future work}

While this manuscript focuses on laying out the fluxon methodology and coronal modeling framework, ongoing work will expand consideration to a further number of Carrington rotations throughout the solar cycle, with a followup manuscript in preparation. \edit1{In particular a higher cadence of rotations allows for condensation of data into time-latitude maps to follow the ebb and flow in time and space of modeled solar wind outflows.}

Other code enhancements in progress involve more complex / faster solar wind solutions, mass loading along fluxons, fluid pressure forces, and variable magnetic flux \edit1{for each fluxon}. \edit1{For FLUX model solutions out to 21.5~R$_\odot$ (0.1~AU), the effects of the Parker Spiral may become an issue when linking open fieldlines down to their photospheric source. A Parker Spiral radially-dependent pseudoforce can be added to the set of force laws to consider this effect on mapping connectivity in more detail.}

\begin{acknowledgements}
The authors gratefully acknowledge support of this work through NASA Heliophysics Living with a Star program grant 80NSSC17K0683, \edit1{AFOSR grant FA9550-19-S-0003, and NASA Heliophysics Supporting Research grant NNH20ZDA001N.} HMI Magnetic field maps are procured courtesy of NASA/SDO and the HMI science team. \edit1{The authors also are particularly thankful for the insightful and helpful referee comments and suggestions.}
\end{acknowledgements}

\software{SunPy \citep{stuart_j_mumford_2023_7850372, sunpy_community2020}, Astropy \citep{astropy:2013, astropy:2018, astropy:2022}}

\bibliography{fluxon}

\listofchanges

\end{document}